\documentclass[aps,prl,twocolumn,superscriptaddress,noshowpacs,
noshowkeys]{revtex4-1}
\usepackage{amssymb, amsmath}
\usepackage{color}
\usepackage{graphicx}

\begin{document}
\title{Tunable mass separation via negative mobility} 
\author{A. S{\l}apik}
\affiliation{Institute of Physics and Silesian Center for Education and Interdisciplinary Research, University of Silesia, 41-500 Chorz{\'o}w, Poland}
\author{J. {\L}uczka}
\affiliation{Institute of Physics and Silesian Center for Education and Interdisciplinary Research, University of Silesia, 41-500 Chorz{\'o}w, Poland}
\author{P. H\"anggi}
\affiliation{Institute of Physics, University of Augsburg, D-86135 Augsburg, Germany}
\affiliation{Nanosystems Initiative Munich, Schellingstr. 4, D-80799 M\"unchen, Germany}
\author{J. Spiechowicz}
\affiliation{Institute of Physics and Silesian Center for Education and Interdisciplinary Research, University of Silesia, 41-500 Chorz{\'o}w, Poland}
\affiliation{Institute of Physics, University of Augsburg, D-86135 Augsburg, Germany}
%
\begin{abstract}
A prerequisite for isolating diseased cells requires a mechanism for effective mass-based separation. This objective, however, is generally rather challenging because typically no valid correlation exists between the size of the particles and their mass value. We consider an inertial Brownian particle moving in a symmetric periodic potential and subjected to an externally applied unbiased harmonic driving in combination with a constant applied bias. In doing so we identify a most efficient separation scheme which is based on the anomalous transport feature of negative mobility, meaning that the immersed particles move in the direction opposite to the acting bias. This work is  first of its kind in demonstrating a {\it tunable} separation mechanism in which the particle mass targeted for isolation is  effectively controlled over a regime of nearly two orders of mass-magnitude upon changing solely the frequency of the external harmonic driving.
This approach may provide mass selectivity required in present and future separation of a diversity of nano and micro-sized particles of either biological or synthetic origin.
\end{abstract}
\maketitle
The objective of separating and sorting  particles of small size is attracting growing interest  \cite{hanggi2009,denisov2014,Reimann_PR_2002,skaug2018, schwemmer2018, dicarlo2012, hanggi2012, eichhorn2012, reimann2012, rubi2014, eichhorn2017, yeomans2017, sens2018, schmelcher2018, zhang2013, heinke2016, han2013, nguyen2014, lim2016, li2018}, opening the way towards the precise analysis of biophysical and synthetic processes on the microscale. In this regime of sizes the omnipresent Brownian jitter dynamics is of relevant impact. Nowadays effective isolation and separation techniques are of essential importance in a wide range of areas including both research and industrial applications. 
Particularly, such techniques carry a large potential  to study selective transport of biological particles such as whole  cells, organelles or DNA complexes. It has been found that several diseases alter physical properties of cells and therefore their sorting has great significance in health care \cite{suresh2007}. So far much emphasis has been placed on size based isolation techniques \cite{dicarlo2012,hanggi2012,rubi2014,eichhorn2017, yeomans2017,sens2018,zhang2013,heinke2016,han2013,nguyen2014,li2018}.
However, another aspect representing one of the most important factors for specifically identifying a bioparticle presents its own mass. For instance, cancer cells are found to differ in mass as compared to healthy ones \cite{suresh2007}. This fact suggests that mass heterogeneity might be an important factor associated with disease initiation and progression. A reliable and effective approach to separating particles by their masses is therefore much in demand and hence one needs to learn more about various mechanisms for separating  different masses on the Brownian scale \cite{schmelcher2018,hanggi1999,marchesoni2007}. As is commonly appreciated, the task is challenging because of the fact that heavier objects do not necessarily imply also larger sizes. This very feature thereby  excludes  passive mechanical separation techniques such as  filtration in artificial sieves \cite{sen2014}.

In the following we demonstrate a nonintuitive, yet efficient, mass-based separation strategy taking advantage of a paradoxical mechanism of negative mobility (NM) \cite{Eichhorn2002,machura2007,speer2007,nagel2008,spiechowicz2014pre,denisov2014}. In a regime of NM the particles move in a direction opposite to the net acting force. 
This phenomenon rests on two main ingredients: (i) a spatially periodic nonlinear structure together with  (ii) an inertial nonequilibrium stochastic dynamics created for example via a time-periodic varying driving force of vanishing mean value. We demonstrate that under an additional action of an applied  constant bias only particles of a given mass migrate in the direction opposite to this net force whereas the others move concurrently towards it. This opens the possibility of steering different particle species in opposite directions under identical experimental conditions. Moreover, we demonstrate that the mass targeted for separation can be tuned by nearly two orders of magnitude by changing only the frequency of external time-periodic driving. The proof of principle experiment of a similar separation scheme but based rather on the particle size has been already demonstrated in Refs. \cite{ros2005, eichhorn2010}, using a lab-on-a-chip device, consisting of  insulator dielectrophoresis in a nonlinear, symmetric microfluidic structure with electrokinetically induced transport. This system was built employing a photolitographic device fabrication strategy without the need of making use of more complex nanofabrication techniques. Very recently, it allowed to induce NM not only for colloidal particles but even for a biological compound in the form of mouse liver mitochondrium \cite{luo2016}. Therefore the separation scheme proposed here may provide mass selectivity required for individual isolation of nano- and micro-particles, proteins, organelles and cells. Yet other suitable setups which allow to test our theoretical predictions would be based on cold atoms dwelling optical lattices \cite{renzoni2003, lutz2013}.

Let us consider a classical inertial Brownian particle dynamics of mass $M$ moving in a spatially periodic one-dimensional potential $U(x) = U(x + L)$ of period $L$ which is subjected to an unbiased time-periodic force $A\cos{(\Omega t)}$ of amplitude $A$ and angular frequency $\Omega$, as well as an external static force $F$. The Brownian dynamics of such a particle is described by the Langevin equation \cite{slapik2018}
\begin{equation}
	\label{model}
	M\ddot{x} + \Gamma\dot{x} = -U'(x) + A\cos{(\Omega t)} + F + \sqrt{2\Gamma k_B T}\,\xi(t).
\end{equation}
The parameter $\Gamma$ denotes the friction coefficient and $k_B$ is the Boltzmann constant. The periodic potential $U(x)$ is taken to possess  \emph{reflection-symmetry} with period $L$ and a potential  height $2\Delta U$, i.e.
\begin{equation}
	\label{potential}
	U(x) = \Delta U \sin \left(\frac{2\pi}{L}x\right).
\end{equation}
The interaction with a heat bath of temperature $T$ is described by thermal fluctuations
modelled by Gaussian white noise of zero mean and unit intensity, i.e.
\begin{equation}
	\langle \xi(t) \rangle = 0, \quad \langle \xi(t)\xi(s) \rangle = \delta(t-s).
\end{equation}
Despite the apparent simplicity of this model setup it exhibits peculiar transport behaviours including e.g. a nonequilibrium noise enhanced transport efficiency \cite{spiechowicz2014pre}, anomalous diffusion \cite{spiechowicz2017scirep}, amplification of normal diffusion \cite{reimann2001,lindner2016}, or  also a  non-monotonic temperature dependence of normal diffusion \cite{spiechowicz2016njp}.

For our analysis we first recast Eq. (\ref{model}) into its dimensionless form; i.e.,
\begin{equation}
	\label{dimlessmodel}
    m \ddot{\hat{x}} + \dot{\hat{x}} = - \hat{U}'(\hat{x}) + a \cos{(\omega \hat{t})} + f + \sqrt{2 D} \, \hat{\xi}(\hat{t}),
\end{equation}
where $\hat{x} = x/L, \quad \hat{t} = t/\tau_0$ and $\tau_0 = \Gamma L^2/\Delta U$.
The dimensionless mass $m$ is given by the  ratio  of two characteristic timescales, reading
\begin{equation}
	\label{mass}
m = \frac{\tau_1}{\tau_0} = \frac{M\Delta U}{\Gamma^2 L^2},
\end{equation}
where \mbox{$\tau_1 = M/\Gamma$}.
 We here emphasize  the fact that the dimensionless mass $m$ depends not only on the actual physical mass of the particle $M$ but as well also on the friction coefficient $\Gamma$ as well as the parameters of the potential; i.e. on half of its barrier height $\Delta U$ and the period $L$.  The strength of friction $\Gamma$ enters inversely the scaled mass value and it implies that the rescaled mass assumes in the regime of viscous moderate-large friction (i.e. operating at low Reynolds numbers, being typical for microsized particles immersed in solution \cite{purcell,lauga}) a rather small value. The dimensionless noise intensity reads $D = k_BT/\Delta U$. The remaining quantities are explicitly defined as detailed in Ref. \cite{slapik2018}. From here on, we stick throughout to these  dimensionless variables. In order to simplify the notation, we will also omit the \emph{hat} notation in Eq. (\ref{dimlessmodel}).
\begin{figure}[t]
	\centering
	\includegraphics[width=0.88\linewidth]{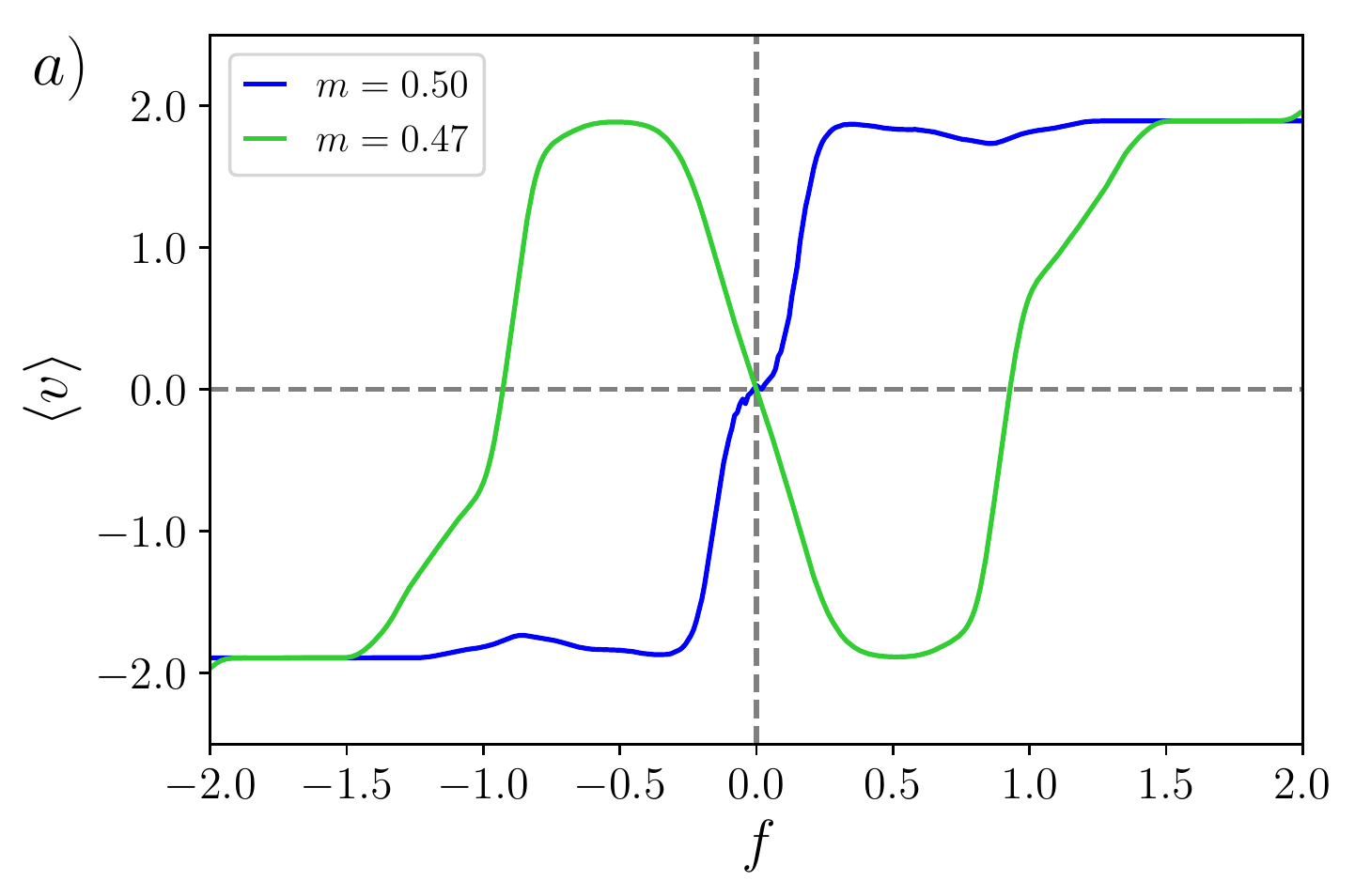}
	\includegraphics[width=0.88\linewidth]{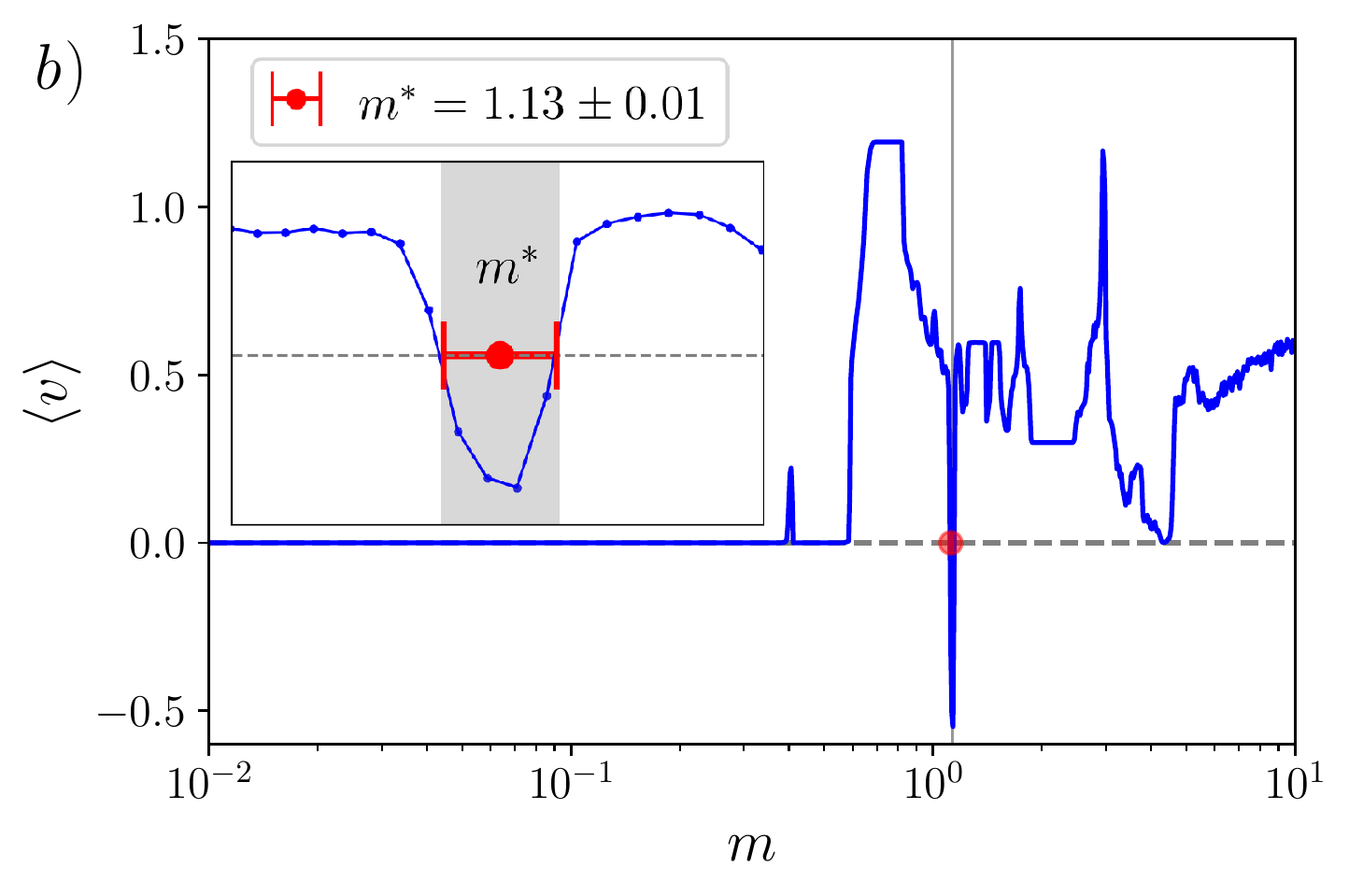}
	\caption{(color online) Panel (a): The force-directed velocity curve $\langle v \rangle(f)$ is depicted for the two parameter regimes corresponding to normal (blue) and anomalous (green) transport behaviour in the form of  NM. Note the sensitivity of the latter effect with respect to changes of the dimensionless mass $m$. The chosen parameters are: $a = 10$, \mbox{$\omega = 5.95$} and $D = 0.001$. Panel (b): the directed velocity $\langle v \rangle$ versus  mass $m$. In the inset we present the blow up (red region) showing the interval of the NM-phenomenon as marked  by the grey area. Parameters are: $a = 5.125$, $\omega = 3.75$, $f = 1$, $D = 0.0001$.}
	\label{fig1}
\end{figure}

The observable of main interest for mass separation is the directed velocity $\langle v \rangle$ of the particle, reading \cite{slapik2018}
\begin{equation}
	\label{dc}
	\langle v \rangle = \lim_{t \to \infty} \frac{1}{t} \int_0^t ds \, \langle \dot{x}(s) \rangle,
\end{equation}
where $\langle \cdot \rangle$ indicates averaging over the thermal noise realizations as well as over the initial conditions for the position $x(0)$ and velocity $\dot{x}(0)$ of the Brownian particle. The latter is required in the deterministic limit $D \propto T \to 0$ when the deterministic dynamics may turn out to be  non-ergodic and dependent on specific choice of these initial conditions \cite{spiechowicz2016scirep}.

Knowingly, the Fokker-Planck equation corresponding to the Langevin Eq. (\ref{dimlessmodel}) cannot be solved analytically in closed form. The task is thus to  systematically analyse by comprehensive numerical means the emerging and rich variety of possible transport behaviors. The setup comprises a complex $5$-dimensional parameter space $\{m, a, \omega, f, D\}$. We nonetheless succeeded in performing the numerical analysis with  unprecedented resolution. 
Overall we considered nearly $10^9$ different parameter sets. The high precision was made possible solely due to  an innovative computational method which is based on employing GPU supercomputers, for details see  in Ref. \cite{spiechowicz2015cpc}.
\begin{figure}[t]
	\centering
	\includegraphics[width=0.9\linewidth]{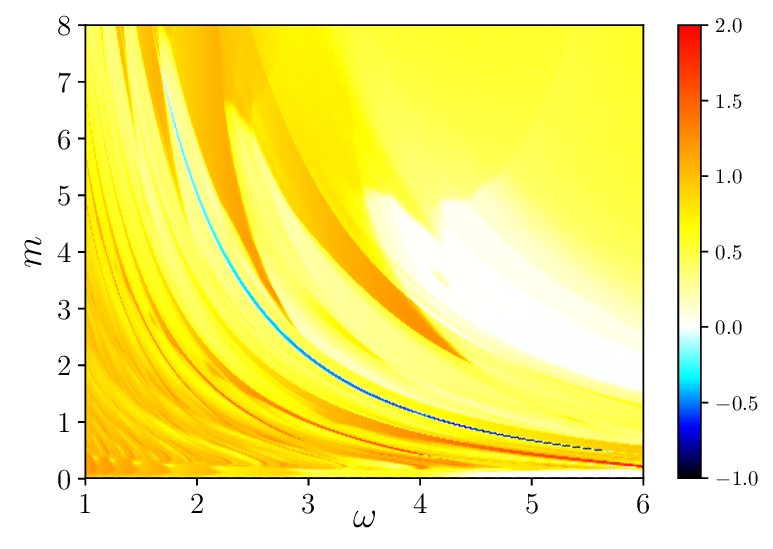}
	\caption{(color online) Two dimensional map of the directed velocity $\langle v \rangle$ of the Brownian particle as a function of the externally applied periodic  driving frequencies $\omega$ versus mass $m$. The magnitude of velocity $\langle v \rangle$ is indicated by the colour code with blue indicating regimes with  NM. Chosen remaining parameters are set at $a = 5.9375$, $f = 1$, $D = 0.0001$.}
	\label{fig2}
\end{figure}

The underlying  symmetries of the  Langevin dynamics in Eq. (\ref{dimlessmodel}) imply that the directed velocity $\langle v \rangle$ behaves odd as a function of the external static bias $f$; i.e.,  $\langle v \rangle(-f) = -\langle v \rangle(f)$ so that $\langle v \rangle(f = 0) \equiv 0$ \cite{denisov2014}. 
Generally, $\langle v \rangle$ is an increasing function in the direction of the static bias $f$ as one commonly would expect. The resulting  particle transport velocity thus follows in the direction of the acting bias $f$, 
$\langle v \rangle =  \mu(f) f$, with a positive-valued nonlinear mobility $\mu(f)> 0$. However, in the parameter space there occur also regimes for which the particle moves on average in the  opposite direction to the applied bias; i.e. $\langle v \rangle < 0$ for $f > 0$ thus exhibiting anomalous transport  in the form of NM with  $\mu(f) < 0$ \cite{machura2007,speer2007,nagel2008,spiechowicz2014pre}. The key prerequisite for the occurrence of the latter phenomenon is that the system (i) is driven far out of thermal equilibrium into a time-dependent  {\it nonequilibrium} state, whose inertial dynamics  does   exhibit (ii)  zero-crossings of $\langle v \rangle$ \cite{machura2007,speer2007}. In our case this condition is induced  by the presence of the external time-periodic driving $a\cos{(\omega t)}$ which in turn {\it overrides}  the limiting response behavior encoded with the  Le-Chat\'elier-Braun equilibrium principle \cite{landau}, stating  that at finite  $f$ the response occurs into the direction of the applied  force towards a new, displaced  equilibrium.
\begin{figure}[t]
	\centering
	\includegraphics[width=0.88\linewidth]{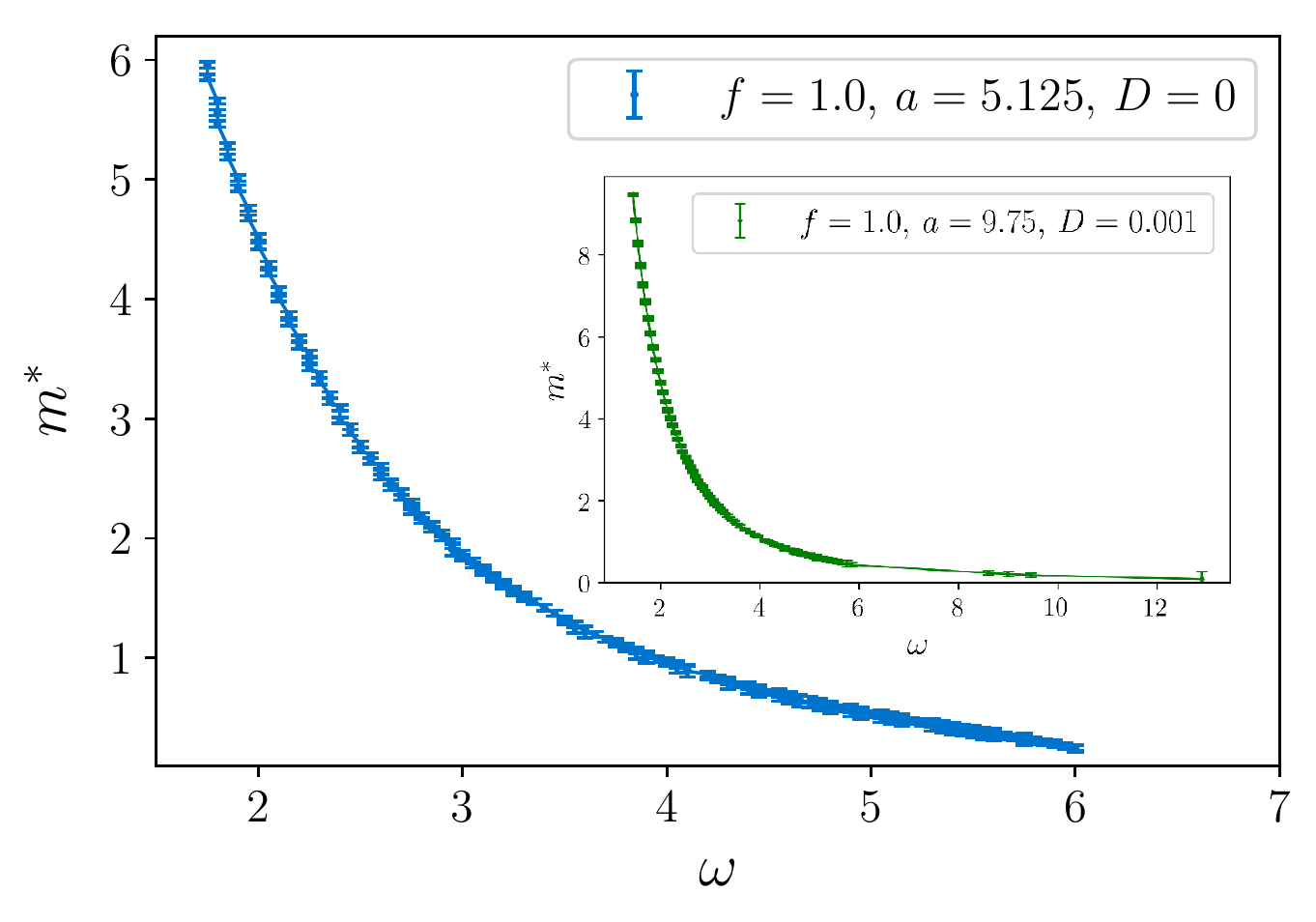}
	\caption{(color online) The dependence of the mass $m^*$ tailored for separation as a function of the external driving frequency $\omega$ for zero temperature $D \propto T = 0$ with fixed $a = 5.125$ and $f = 1$. In the inset we present a different regime with $D = 0.001$ allowing for efficient and tunable mass separation via negative mobility.}
	\label{fig3}
\end{figure}

With  panel (a) of Fig. \ref{fig1} we exemplify two force-velocity characteristics, $\langle v \rangle(f)$, corresponding to normal and NM transport behaviour. Please note the sensitivity of the latter effect with respect to minute changes in the dimensionless mass $m$. A tiny change in mass by $\Delta m = 0.03$ is accompanied with the reverse of the particle response (note the blue line versus the green line behavior).  There is seemingly no clear relationship detectable among the set of  parameter values and the occurrence of this NM phenomenon. A small displacement in the parameter space may either cause a sudden emergence of  NM or its rapid absence. One important observation is that typically it occurs in  regimes for which the values of the parameters are  a priori unknown. The interested  reader is referred to the animation of the maps $\langle v \rangle(\omega, m)$ for several different magnitudes of the ac-driving amplitude $a$ which can be inspected on the web \cite{animation}.

Amongst the regimes of NM in the parameter space there are tailored ones for which the latter phenomenon appears only for a very narrow interval of the mass $m$. This is illustrated with panel (b) of Fig. \ref{fig1} where we depict the directed velocity $\langle v \rangle$ of the Brownian particle versus the mass $m$. An interesting transport property can be detected: amongst many particles with masses from a wide interval $m \in [10^{-2},10^1]$ only those with a mass $m \approx 1.13$ will move in the opposite direction $\langle v \rangle < 0$ to the acting bias $f = 1 $. All other particles with positive velocity $\langle v \rangle$ will follow towards the direction of the bias. As a result, the particles with mass very close to $m \approx 1.13$ will be separated from all the others. This process of mechanical separation seems to be very promising provided that one would be able to control the mass $m^*$ of particles which are intended to be separated. The half-width $\delta m \approx 0.01$ of the interval where the NM occurs is indicated in the inset of the panel (b). It can be viewed as the resolution capacity for separation. However, we stress that typically NM in these intervals is sharply peaked meaning that due to the clear difference between the magnitude of the negative velocity indeed only the particles with the precisely defined mass $m^*$ will be pronouncedly isolated from the others moving in the same or the opposite direction. We in turn undertook the attempt to search for such parameter regimes of the Langevin equation (\ref{dimlessmodel}) that would enable to control the occurrence of the NM by tuning just one parameter.
\begin{figure}[t]
	\centering
	\includegraphics[width=1.0\linewidth]{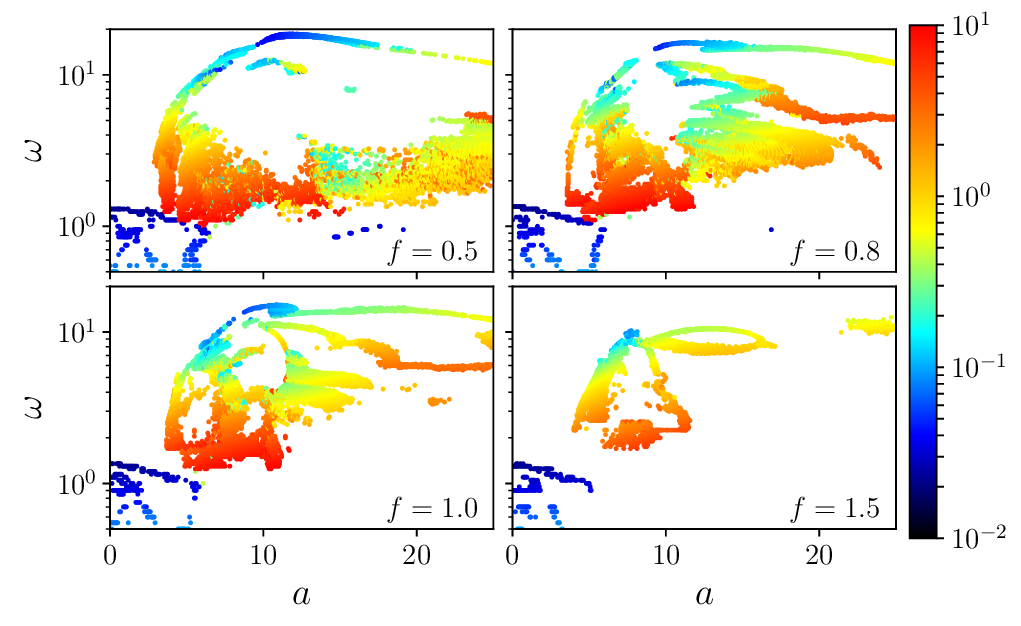}
	\caption{(color online) The mass $m^*$ targeted for separation (color coded scale) via the NM effect as a function of the external ac-driving strength $a$ and frequency $\omega$ for different values of the bias $f$. Thermal noise intensity is set to zero $D = 0$.}
	\label{fig4}
\end{figure}

After this comprehensive numerical analysis we identified practically  all sets of parameters $\{a, \omega, f, D\}$ for which  NM emerges only for a narrow interval of mass $[m^*-\delta m,m^* + \delta m]$. Among selected parameter regimes, we focused on those which reveal a specific functional dependence between the isolated mass $m^*$ and the parameter $a$, $\omega$, $f$ or $D$. In Fig. \ref{fig2} we present such an example of the directed velocity $\langle v \rangle$ map as a function of the external driving frequency $\omega$ and scaled mass $m$. The magnitude of the velocity $\langle v \rangle$ is indicated with a corresponding colour code. The blue regime corresponds to NM. It can be observed that for a given value of the frequency $\omega$  NM is present {\it solely} for a particular value of mass $m^*$. Therefore, using those tailored parameters read off from Fig. \ref{fig2} we are able to tune the NM to the particle of a given mass $m^*$ by changing the value of the external driving frequency $\omega$ at fixed periodic driving strength. Doing so allows for  efficient mass separation from an interval covering nearly two orders of magnitude.

In order to examine our results even more accurately we isolated the NM area from corresponding two parameter maps into a graph of the target mass $m^*$ versus the external driving frequency $\omega$.
The result is depicted in Fig. \ref{fig3}  for zero temperature $D \propto T = 0$. 
A change of thermal noise intensity from $D=0$ up to temperature $D \approx 0.0003$ does not significantly alter the desired characteristics which is quite robust with respect to temperature variation.  Because NM derives from the complex deterministic dynamics, strong thermal noise of sufficient intensity is expected to cause a blurring of the phenomenon \cite{slapik2018}. Interestingly, however, increasing thermal noise strength produces a shrinking of corresponding  NM intervals, thereby optimizing the range $\delta m$. This feature implies an improvement of the selectivity for separation. In the inset of the panel (a) we additionally depict different parameter regime allowing tunable mass separation for even higher temperature $D = 0.001$.  
Moreover, our method of mass separation is stable against a variation of the amplitude strength $a$ (not depicted). The effect is present for a wide range of amplitudes $a \in [4,8]$. 
At this point we remark that the effect of mass separation upon harvesting the NM phenomenon is also present for the dependence $m^*(a)$ and alike for $m^*(f)$; the range of tunable mass separation proves, however, somewhat smaller.

Finally, we consider yet a further issue: Let us assume that we deal with a given mass $m^*$ which we want to separate from the rest. The question then is: for how many different masses $m^*$ taken from the extended  interval $ m^* \in  [10^{-2}, 10^1]$ is it possible to isolate the sought-after parameter set $\{a, \omega, f, D\}$ for which only this very specific mass value $m^*$ displays  NM, thereby allowing its separation in a most efficient unique manner. We note that for a single mass $m^*$ there might be several parameter regimes obeying this condition. The answer to this question is summarized with Fig. \ref{fig4}. There, the distribution of the mass $m^*$ targeted for separation via the NM phenomenon in the parameter plane of the ac-driving amplitude $a$ and frequency $\omega$ is depicted for different values of the static bias $f$. We observe that small masses can be isolated with low values of $a$ and $\omega$. Medium and large masses are separated when the amplitude and frequency assume moderate magnitudes. We detected that with this method and for a fixed bias value $f = 1$ nearly all masses from the considered interval can be isolated.
We also find that the overall distribution of the mass $m^*$ targeted for separation depicted (in color) in Fig. \ref{fig4} is robust with respect to a variation of bias $f$, noting that for smaller values of $f$ it undergoes a stretching.

In conclusion, this work provides an effective solution for the objective of the tunable mass separation. In our scheme mass targeted for isolation can be controlled by nearly two orders of magnitude by merely changing the frequency $\omega$ of the external harmonic driving. This task apparently cannot be accomplished with similar quality by use of alternative methods such as filtration techniques or schemes which are based on fluid-driven Brownian motor methodology \cite{hanggi2009, kettner2000, matthias2003}. The approach presented here uses only a spatially periodic nonlinear structure in combination with unbiased external time-periodic driving. Our method can further be adapted to the needs by proper fabrication of the nonlinear landscape described by its barrier height $\Delta U$ and period $L$. Other advantages are: (i) as a representative of active techniques it offers an improved averaged migration speed as compared to alternative approaches \cite{luo2016}, (ii) in contrast to microfluidic methods it allows the possibility to not only deflect different particle species along different transport angles, but even to steer them in opposite directions, (iii) use of small size of a lab-on-a-chip device technology together with advantageous fabrication costs allows for massive parallelization which makes high-throughput separation possible. We envision that the separation strategy proposed here allows for  mass selectivity isolation of nano- and micro-particles, proteins, organelles and cells.

This work was supported by the Grant NCN 2017/26/D/ST2/00543 (J.S.) and NCN 2015/19/B/ST2/02856 (J.{\L}.)

\end{document}